\documentclass[11pt,a4paper]{article}

\usepackage{amsmath,amssymb}
\usepackage{graphicx}
\usepackage{epsf}
\usepackage{epsfig}
\usepackage[figuresright]{rotating}

\newcommand{\be}{\begin{eqnarray}}

\newcommand{\ee}{\end{eqnarray}}

\begin{document}

\title{Some Comments About The High Energy Limit of QCD \\}
\author{Larry McLerran \\
 {\small\it Physics Department and Riken Brookhaven Center}\\
 {\small\it  PO Box 5000, Brookhaven National 
Laboratory,  
 Upton, NY 11973 USA} \\
        }
  \date{}
  \maketitle

\abstract{ I argue that the physics of the scattering of very high energy 
strongly interacting particles is controlled by a new, universal
form of matter, the Color Glass Condensate.  This matter is the 
dominant contribution to the low x part of a hadron wavefunction.  In collisions, this matter 
almost instantaneously turns into a Glasma.
The Glasma initially has strong longitudinal color electric and magnetic fields, with topological charge.
These fields melt into gluons.  Due to instabilities, quantum noise is converted into classical turbulence, which may be responsible for the early thermalization seen in heavy ion collisions at RHIC.}

\section{The High Energy Limit}

The high energy limit of QCD is the limit where the energy of collisions goes to infinity, but the 
typical momentum transfer is finite.  This momentum transfer can be much larger than
$\Lambda_{QCD}$, but it is to  remain fixed.  This is not the short distance limit, where both
momentum transfer and energy go to infinity.  The short distance limit is understood using weak coupling perturbation theory.  The high energy limit is that of non-perturbative phenomena such as Pomerons, Reggeons, unitarization etc. etc.  One of the purposes of this lecture is to convince the reader
that this non-perturbative limit of QCD is also a weak coupling limit

The Bjorken x variable can be understood as the ratio of the energy of the constituent of a hadron 
to that of the hadron itself in the reference frame where the hadron has large energy.  The typical minimal value of x is 
\be
	x_{min} \sim \Lambda_{QCD}/E_{hadron}
\ee
The minimal value decreases as the hadron energy increases.

\begin{figure}[ht]
    \begin{center}
        \includegraphics[width=0.40\textwidth]{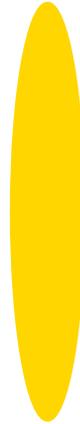}
        \caption{The various componenents of the hadron wavefunction. }
\label{wfn}
    \end{center}
\end{figure}

A hadron wavefunction has many different components.  This is illustrated in Fig. \ref{wfn}.
A nucleon has a Fock space component with three quarks and no gluons, with an extra gluon and with many extra gluons.  The components which control low energy scattering are those with three quarks and a few gluons.  For high energy scattering processes, the typical matrix elements are those with three quarks, many quark anti-quark pairs, and even more gluons.

\section{What is the Color Glass Condensate?}

The original ideas for the Color Glass Condensate were motivated by the 
result for the
HERA data on the gluon distribution function shown in 
\begin{figure}[!htb]
   \begin{center}
       \mbox{{\epsfig{figure=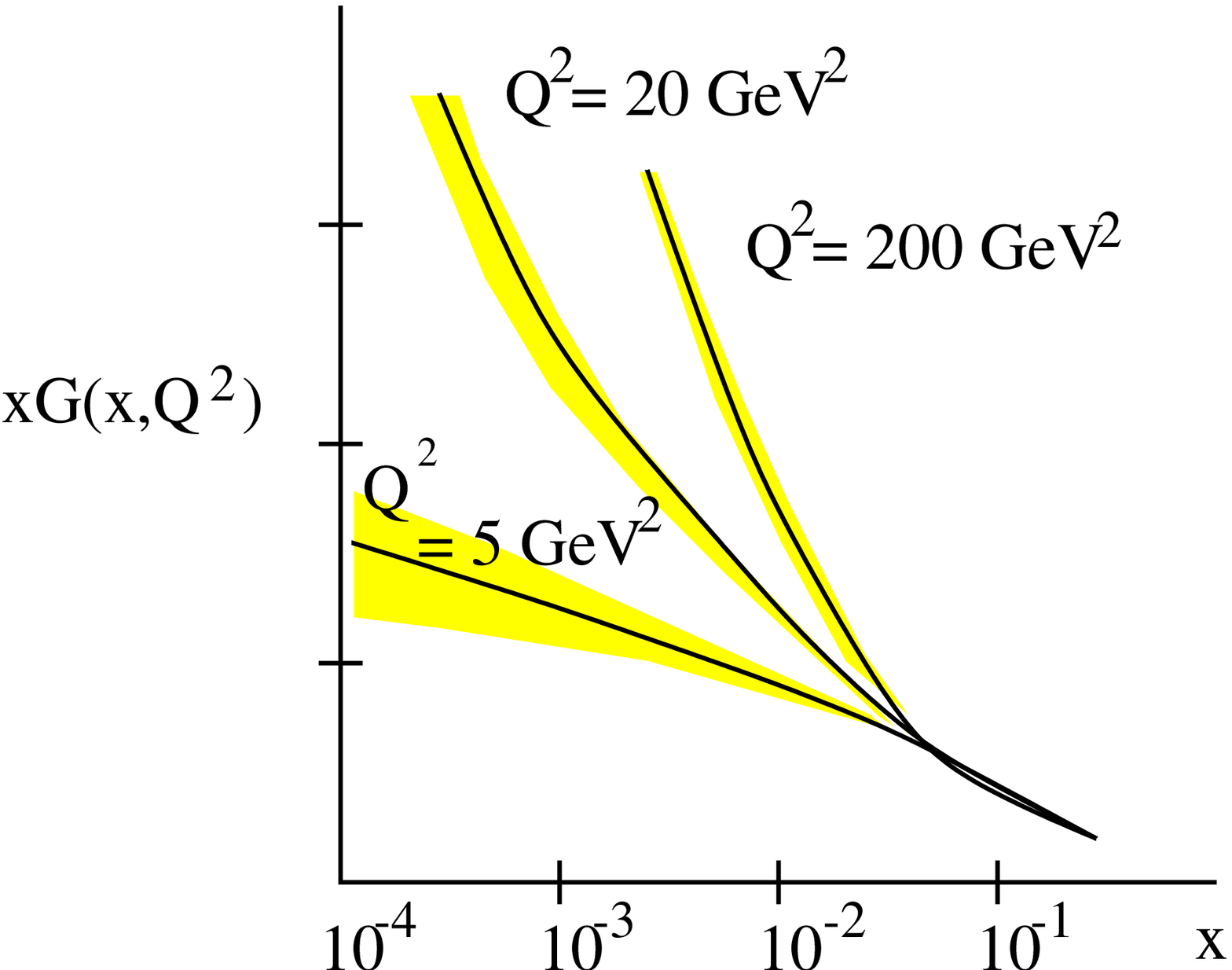,
        width=0.50\textwidth}}\quad
             {\epsfig{figure=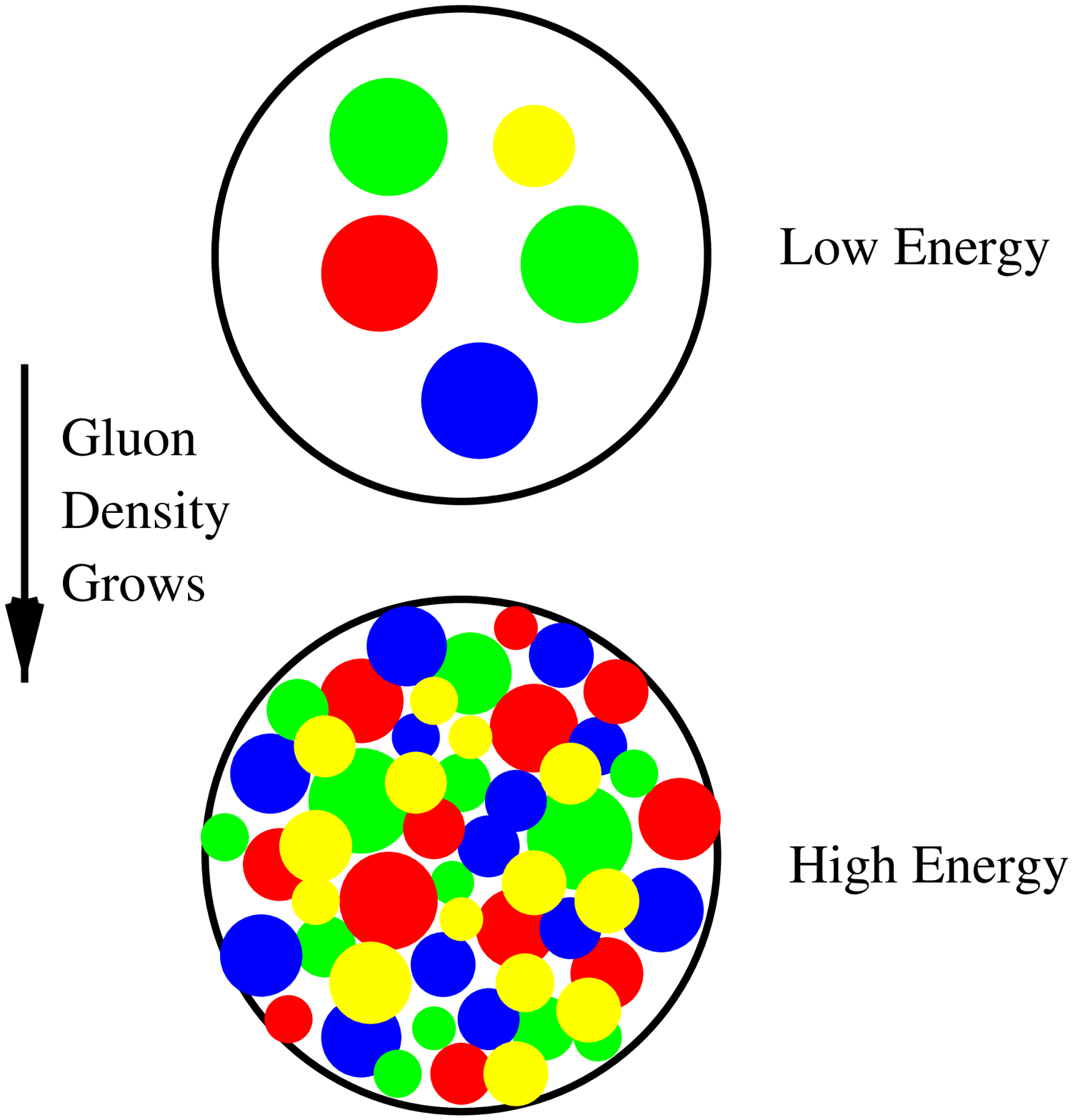,
        width=0.50\textwidth}}}
   \end{center}
\caption[*]{(a)The HERA data for the gluon distribution function
as a function of x for various values of 
{\protect{$Q^2$}}. 
(b) 
A physical picture of the low x gluon density inside a hadron
as a function of energy.}
       \label{heradata}
\end{figure}
Fig. \ref{heradata}(a) \cite{hera}  The gluon density is rising rapidly as a function of
decreasing x.  This was expected in a variety of 
theoretical works\cite{glr}-\cite{bfkl} and has the implication that the 
real physical transverse density of gluons must 
increase.\cite{glr}-\cite{mq},\cite{mv}.  This follows because total cross 
sections rise slowly at high energies but the number of gluons is rising 
rapidly.  The gluons must fit inside the size of the hadron.

This is shown in Fig. \ref{heradata}(b). This led to
the conjecture that the density of gluons should become limited,
that is, there is gluon saturation. \cite{glr}-\cite{mq}, \cite{mv}
Actually, I argue that as one goes to higher energy, a hadron becomes a tightly packed system
of gluons larger than some size scale.  For smaller gluons there are holes.  As one increases
the energy, one still adds in more gluons, but these gluons are small enough that they fit into the holes.
Because in quantum mechanics,  we interpret size as wavelength as inversely proportional to momentum,
at high energies, the gluons are tightly packed for gluons below some momentum, and are filling
in above that momentum.  There is therefore a critical momentum, the saturation momentum,
which characterizes the filling.  This saturation momentum increases as the energy increases, so the total number of gluons can increase without bound.

The low x gluons therefore are closely packed together, and become more
closely packed as the energy increases. The 
strong interaction 
strength must become weak, $\alpha_S \ll 1$.  Weakly coupled systems 
should be possible to understand from first principles in 
$QCD$.\cite{mv}-\cite{ilm}  

This weakly coupled system is called a Color Glass Condensate  (CGC) for reasons 
we now enumerate:\cite{ilm}
\begin{itemize}
\item {\bf Color}  The gluons which make up this matter are colored.
\item{\bf Glass} The gluons at small x are generated from gluons at larger
values of x.  In the infinite momentum frame, these larger momentum gluons
travel very fast and their natural time scales are Lorentz time dilated.  
This time dilated scale is transferred to the low x degrees of freedom
which therefore evolve very slowly compared to natural time scales.  This
is the property of a glass.
\item{\bf Condensate} The phase space density 
\be
\rho = {1 \over {\pi R^2}}{{dN} \over {dyd^2p_T}}
\ee
is generated by a trade off between a negative mass-squared term linear in 
the density which generates the instability, $-\rho$ and an interaction
term $\alpha_S \rho^2$ which stabilizes the system at a phase space density 
$\rho \sim 1/\alpha_S$.  
Because $\alpha_S << 1$,
this means that the quantum mechanical states of the 
system associated with the condensate are multiply occupied.  They are 
highly coherent, and share some properties of Bose condensates.
The gluon occupation factor is very high, of order $1/\alpha_S$, but it is
only slowly (logarithmically) increasing when further increasing the
energy, or decreasing the transverse momentum. This provides saturation
and cures the infrared problem of the traditional BFKL approach.\cite{im2001}

\end{itemize}

One can understand the high phase space occupancy $1/\alpha_S$ from simple arguments.
The momentum scale in the phase space distribution is the De Broglie wavelength of the gluons
which we can interpret as the size of the gluons.  The gluons of fixed size will densely occupy the system
until there  are $1/\alpha_S$ gluons of fixed size closely packing the system.  The gluons
interact with strength $\alpha_S$, so that when $1/\alpha_S$ sit on top of one another, they act coherently like a hard sphere with interaction strength of order 1.

Implicit in this definition is a concept of fast gluons which act as sources 
for the colored fields at small x.  These degrees of freedom are treated 
differently than  the fast gluons which are taken to be sources.  
The slow ones are fields.  There is
an arbitrary $X_0$ which separates these degrees of freedom.  This 
arbitrariness is cured by a renormalization 
group equation which requires that physics be independent of
$X_0$.  In fact this equation determines
much of the structure of the resulting theory as its solution flows to
a universal fixed point.\cite{ilm}-\cite{jklw}

\begin{figure}[ht]
    \begin{center}
        \includegraphics[width=0.40\textwidth]{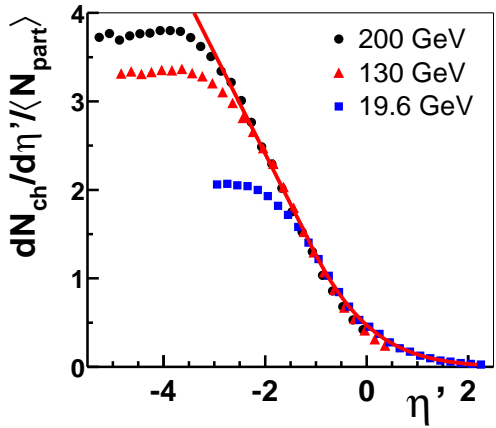}
        \caption{Limiting fragmentation as measured in the Phobos experiment at RHIC. }
\label{limfrag}
    \end{center}
\end{figure}

There is evidence which supports this picture.  One piece
is the observation of limiting fragmentation.
This phenomena is that if particles collide at some fixed 
center of mass energy
and the distribution of particles are measured as a function of their 
longitudinal momentum
from the longitudinal momentum of one of the colliding particles, 
then these distributions do not change as one goes to 
higher energy, except for  the new degrees of freedom that appear.
This is true
near zero longitudinal momentum in the center of mass frame because
new degrees of freedom appear as the center of mass energy is increased.  
In the analogy with the CGC,
the degrees of freedom, save the new ones added in at low longitudinal 
momentum, are the sources.  The fields correspond to the new degrees of
freedom.   The sources are fixed in accord with limiting fragmentation.
One generates an effective theory for the low longitudinal
momentum degrees of freedom as fixed sources above some cutoff, and
the fields generated by these sources below the cutoff.  A recent
measurement of limiting fragmentation comes from the Phobos
experiment at RHIC shown in Fig. \ref{limfrag} \cite{phoboslfrag}

Of course the perfect scaling of the limiting fragmentation curves is
only an approximation.  As shown by Jalilian-Marian, 
the limiting fragmentation
curves are given by the total quark, antiquark and gluon distribution 
functions of the fast particle measured at a momentum scale $Q_{sat}^2$
appropriate for  the particle that it collides with.\cite{jamal}  
The saturation momentum $Q_{sat}$
will play a crucial role in our later discussion.  It is a momentum scale
which is determined by the density of gluons in the CGC
\be
	{1 \over {\pi R^2}} {{dN} \over {dy}} \sim {1 \over \alpha_S} Q_{sat}^2
\ee
The saturation momentum
turns out to depend on the total beam energy because
the longitudinal momentum scale of the target particle at fixed $x$
of the projectile will depend upon the beam energy.  It is nevertheless
remarkable how small these violations appear to be.

The CGC may be defined mathematically by a path integral:
\be
	Z = \int_{X_0} [dA][dj] exp\left(iS[A,j] - \chi[j] \right)
\ee
What this means is that there is an effective theory defined below some cutoff 
in $x$ at $X_0$, and that this effective theory is a gluon field in the 
presence of an external source $j$. This source arises from the quarks
and gluons with $x \ge X_0$, and is a variable of integration.
The fluctuations in $j$
are controlled by the weight function $\chi[j]$.  It is $\chi[j]$
which satisfies renormalization group equations which make the theory
independent of $X_0$.\cite{ilm},\cite{jkmw}-\cite{jklw},\cite{b}-\cite{jimwlk}.  The equation for
$\chi$ is called the JIMWLK equation.  This equation reduces in 
appropriate limits to the BFKL and DGLAP evolution
equations.\cite{bfkl}, \cite{dglap} 
The theory above is mathematically very similar to that of spin glasses.

There are a variety of kinematic regions where one can find
solutions of the renormalization group equations which have
different properties.  There is a region where the gluon density
is very high, and the physics is controlled by the CGC.  This is when
typical momenta are less than a saturation momenta which depends on $x$,
\be
	Q^2 \le Q_{sat}^2(x)
\ee
The dependence of $x$ has been evaluated by several authors,
\cite{glr},\cite{lt}-\cite{mt}, 
and in the energy range appropriate for current experiments
has been determined by Triantafyllopoulos to be 
\be
	Q_{sat}^2 \sim (x_0/x)^\lambda~GeV^2
\ee
where  $\lambda \sim 0.3$.  The value of $x_0$
is not determined from the renormalization group equations and must be 
found from experiment.

\begin{figure}[ht]
    \begin{center}
        \includegraphics[width=0.40\textwidth]{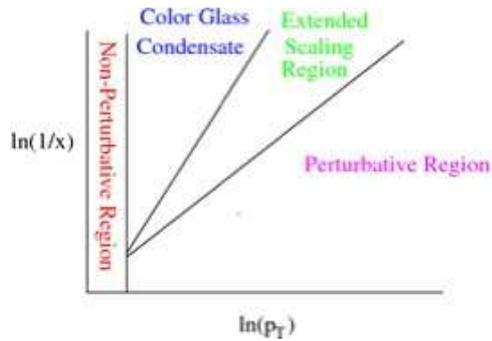}
        \caption{The kinematic regions of the Color Glass Condensate. }
\label{xq2}
    \end{center}
\end{figure}

The kinematic region corresponding to the CGC is shown in Fig. \ref{xq2}. 

There is also a region of very high $Q^2$ at fixed x, where the density of
gluons is small and perturbative QCD is reliable.  It turns out there is
a third region intermediate between high density and low where there are
universal solutions to the renormalization group equations and scaling
in terms of $Q_{sat}^2$.\cite{iim}  In this region and in the region
of the CGC, distribution functions are universal functions of only 
$Q^2/Q_{sat}^2(x)$.  The extended scaling region is when
\be
	Q_{sat}^2 \le Q^2 \le Q_{sat}^4/\Lambda_{QCD}^2
\ee

\section{What is the Form of the Color Glass Fields?}

\begin{figure}[ht]
    \begin{center}
        \includegraphics[width=0.30\textwidth]{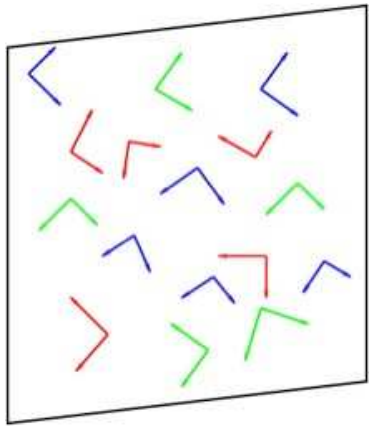}
        \caption{The color glass field. }
\label{sheet}
    \end{center}
\end{figure}

One can simply compute the form of the Color Glass fields.  If we work in a frame
where the hadron has a large momentum, the $z-t \sim 0$.  The only big component
of $F^{\mu \nu}$ is $F^{i+}$ where 
\be
	x^\pm = (z \pm t)/\sqrt{2}
\ee
If we set $F_{i-} = 0$, then simple algebra tells us that the big field strengths	
are $E$ and $B$, and that
\be
	 \vec{E} \perp \vec{B} \perp \vec{z} 
\ee
The fields are plane polarized perpendicular to the beam direction.
These are the Lienard-Wiechart potentials which correspond to a Lorentz boosted Coulmbg field.
They exist within the Lorentz contracted sheet and have a longitudinal extent corresponding
to the fast moving sources.  (The vector potential corresponding to these field is extended, and the wee gluons corresponding to these fields are extended over a larger longitudinal size scale).  The fields have
random polarizations and colors.
This is shown in Fig. \ref{sheet}.

\section{What is the CGC Good For?}

The CGC provides a unified description of deep inelastic structure 
functions, of deep inelastic diffraction and of hadron-hadron collisions 
at high energies.  It is the high energy limit of QCD.  As such, it has 
many tests to pass before being accepted as a correct description.  Over 
the last several years, there have been many qualitative and 
semi-quantitative successes of this description.  It also provides an 
intuitively plausible and mathematically consistent description of such 
phenomena.

One can compute the $x$ dependence of the saturation momentum.\cite{iim}-\cite{mt}
This solutions results from renormalization group equations.  The results agree with Hera
phenomenology.\cite{gbw}  In particular, within the same description one can compute both
deep inelastic structure functions and diffractive structure functions.  

\begin{figure}[ht]
    \begin{center}
        \includegraphics[width=0.65\textwidth]{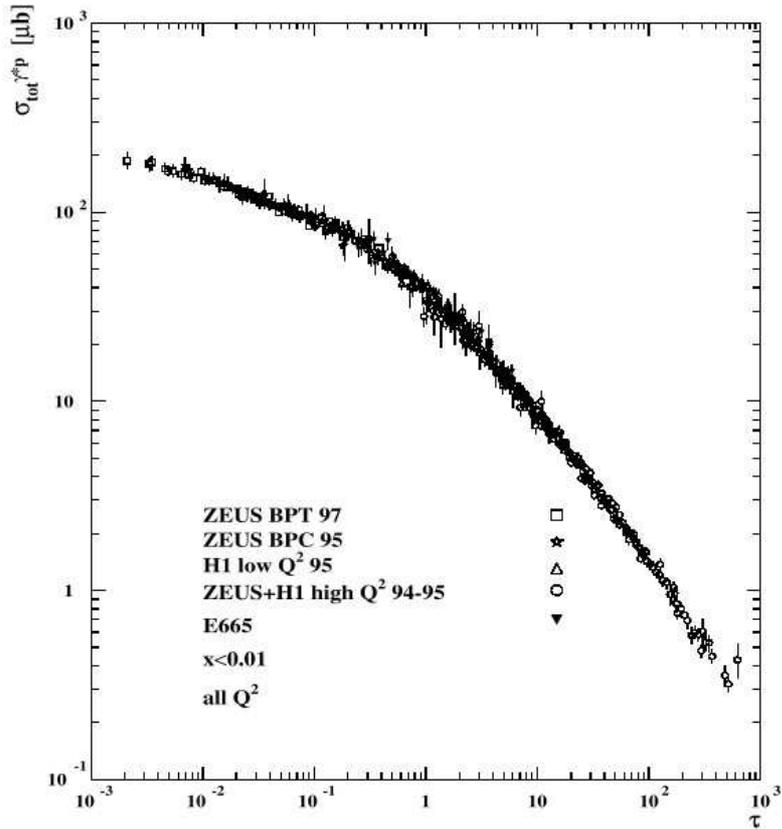}
        \caption{Geometric Scaling as seen at Hera.. }
\label{gscale}
    \end{center}
\end{figure}

In addition, the CGC predicts the existence of geometric scaling.\cite{gbks}  Geometric scaling
means that the cross section for deep inelastic scattering of a virtual photon from a hadron depends only upon the scale invariant ratio $Q^2/Q_{sat}^2$, and not independently $Q^2$ and $x$.  Such 
scaling is shown in Fig. \ref{gscale} for x values of $x < 10^-2$.  

\begin{figure}[ht]
    \begin{center}
        \includegraphics[width=0.45\textwidth]{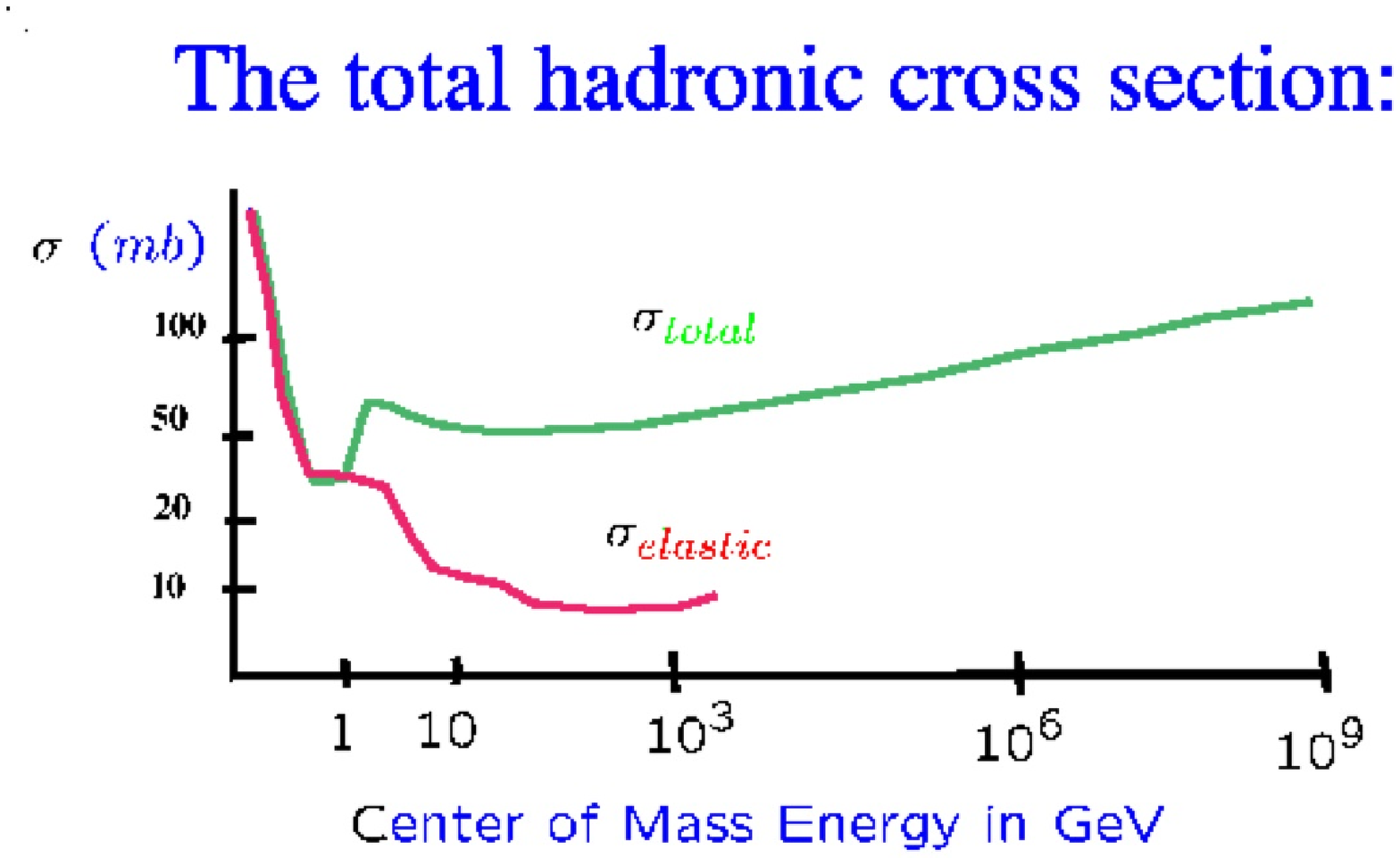}
        \caption{The total hadronic cross section as a function of energy.. }
\label{cross}
    \end{center}
\end{figure}

The total cross section for hadron-hadron scattering is a slowly varying function of energy
as shown in Fig. \ref{cross}. 
The Color Glass Condensate provides a heuristic explanation of this.\cite{wiedeman}-\cite{ikeda}
We assume the distribution of gluons in the transverse plane of a hadron as a function of energy factorizes,
\be
	{{dN} \over {d^2r_T dx}} \sim (x_0/x)^\lambda e^{-2 m_\pi r_T}
\ee
where the exponential fall off should be  of the form shown at large $r_T$, since the lowest mass particle exchange with isospin zero is two pions.  The cross section for a particle of some size to penetrate
the hadron, and have a large probability to scatter occurs when this density is some fixed number.
Therefore
\be
	\sigma \sim R^2 \sim ln^2(1/x) \sim ln^2(E/E_0)
\ee
This behaviour is the Froissart bound for cross sections, and describes high energy scattering
reasonably well.  It origin is the trade off in rapidly falling impact parameter profile against rapidly rising density of partons.

\begin{figure}[ht]
    \begin{center}
        \includegraphics[width=0.70\textwidth]{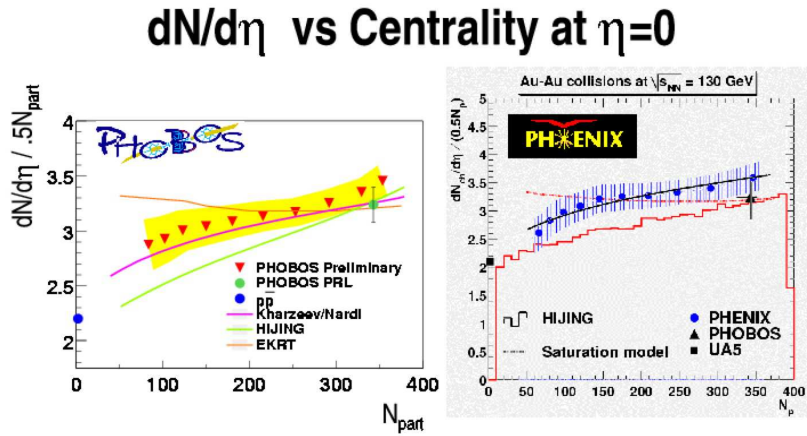}
        \caption{The multiplicity as a function of energy observed at the RHIC }
\label{nard}
    \end{center}
\end{figure}

One of the remarkable predictions of the Color Glass Condensate was the mutliplicty of particles produced in heavy ion collisions at RHIC.\cite{nardi}  The multiplicty of saturated gluons inside a nucleus should scale as
\be
	{{dN} \over {dy}} \sim {1 \over \alpha_S} \pi R^2 Q_{sat}^2
\ee
The energy dependence of $Q_{sat}$ is known, and the dependence on centrality
should be proportional to $N_{coll}$, the number of nucleons colliding, at not too high an energy.
Assuming the hadron multiplicity is proportional to the number of produced gluons gives
the plot shown in Fig. \ref{nard}. \cite{whitepaper}

Experiments  using deuterons on nuclei in the fragmentation region of the deuteron also test ideas about how the gluon distribution is modified by the Color Glass Condensate.  
If one uses a multiple collision model to infer the change of the distribution of produced 
high momentum particles, then there should be more particles at high $p_T$ for intermediate to high
$p_T$.  Because of probability conservation, there should be a depletion of low momentum particles.
At very high $p_T$, where multiple scattering should not be important, the distribution should not be changed.

The CGC on the other hand predicts an additional phenomenon.  Since there is a high density media, the QCD evolution equations stop running when the scale becomes of the order of the saturation momentum.  This means that in nuclei which have a larger saturation momentum than the proton, that there should be a net depletion of particles. 

\begin{figure}[htb]
    \begin{center}
        
\includegraphics[angle=270,width=0.40\textwidth]{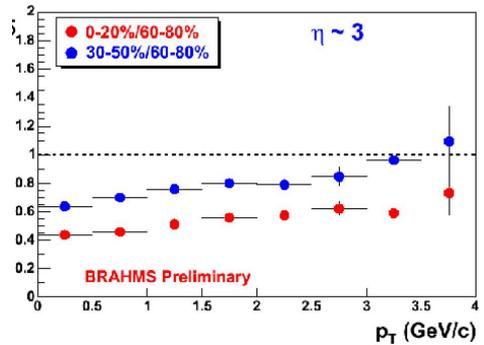}
\caption{The ratio of single particle production in central to peripheral 
collisions at forward rapidity
as a function of centrality as measured in the Brahms experiment.} 
\label{brahms}
    \end{center}
\end{figure}

The experimental measurements
are in accord with CGC predictions at forward rapidities.\cite{whitepaper}  The effect is shown as a function of centrality of the collisions in Fig. \ref{brahms}.  Forward rapidities correspond to small x values for the deuteron wavefunction.  At larger values of x, the multiple scattering effects dominate,
and there is an enhancement as a function of centrality.  For the central region of gold-gold collisions at RHIC energies, the effects almost cancel one another.     

In the future years, there will be increasingly stringent tests arising at 
RHIC, LHC and potentially eRHIC.  Theoretically, we are just beginning to 
understand the properties of this matter.  New ideas concerning the 
structure of the underlying theory and the breadth of phenomena it 
describes are changing the way we think about high energy density matter.

\section{What is the Glasma?}

\begin{figure}[htb]
   \begin{center}
       \includegraphics[width=0.40\textwidth]{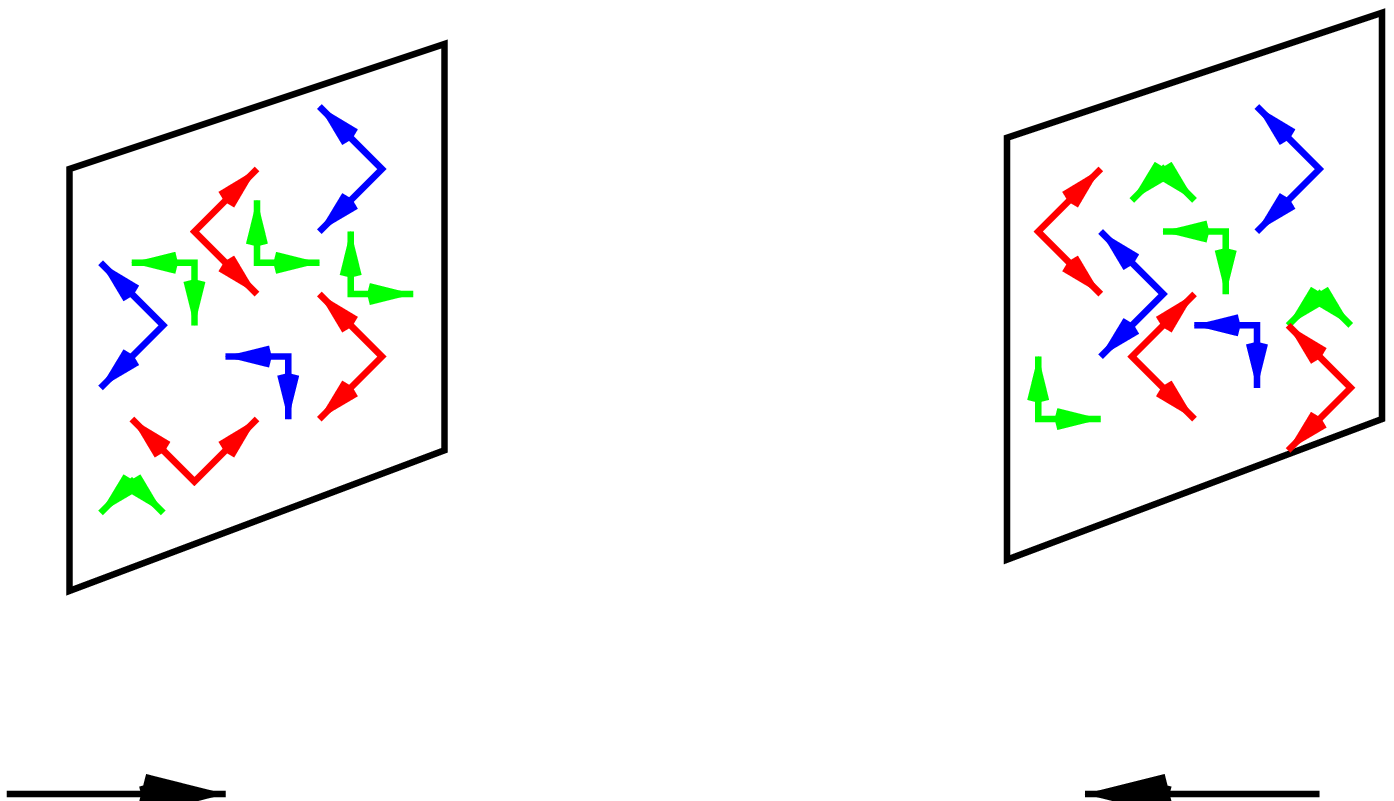}
             \includegraphics[width=0.25\textwidth]{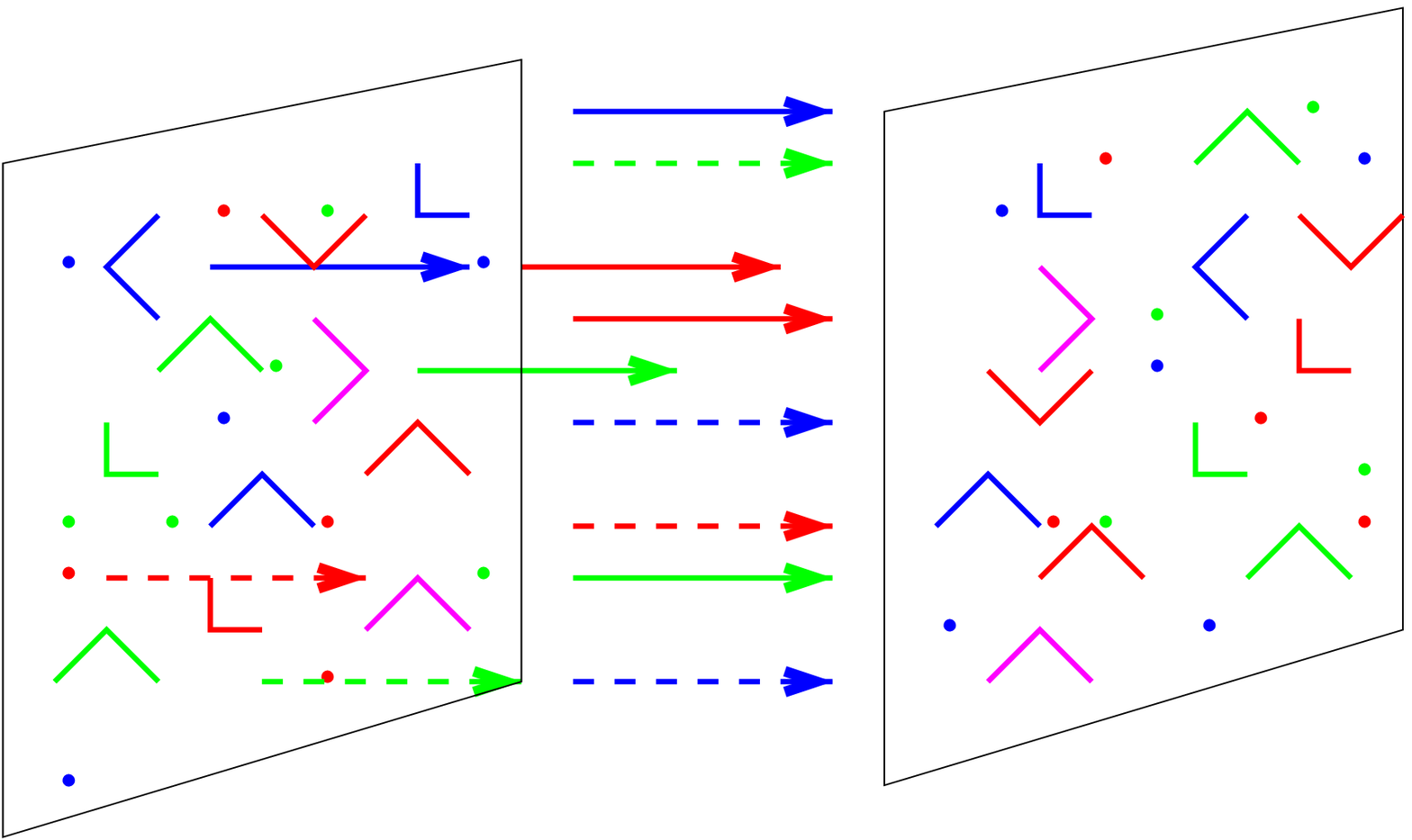}
   \end{center}
\caption[*]{(a)Two sheets of colored glass approach one another. 
(b) 
After the collisions, Glasma is formed in the region between the 
sheets.}
       \label{glasma}
\end{figure}

When two sheets of colored glass collide, the properties of the matter are changed
in the time it take light to propagate across the sheets of colored glass. \cite{mkw}-\cite{lappi}
In Fig. \ref{glasma} a,
the sheets of colored glass approach one another.  The colored fields in the two sheets
form a condensate of Weizsacker-Willams fields disordered in polarization and color.
In the time it takes the sheets to pass through one another, the fast degrees of freedom 
gain a density of color electric and magnetic charge.  The density of charge on each sheet
is equal and opposite.  This is a consequence of the classical fields which are generated by the source 
sheets of color glass.  Attached to the region away from the collision region is a pure two dimensional transverse vector potential.  This potential has no color electric and magnetic field until the nuclei pass through one another, since then the vector potential of one sheet multiplies the field of the other.
Then sources are set up according to the Yang-Mill equations
\be
	\rho_E^a &  = &  f^{abc} A^b \cdot E^c \nonumber \\
	\rho_B^a & = & f^{abc} A^b \cdot B^c 
\ee	 

These colored electric and magnetic charges generate longitudinal color electric
and magnetic fields as shown in Fig. \ref{glasma}b.   The reason why both electric and magnetic
fields are made is because of the duality of the Yang-Mill equations under $E \leftrightarrow B$,
and because the initial fields of the colored glass have this symmetry.

When there is a nonzero $E\cdot B$ means that there is a topological charge induced.
This Chern Simons charge is
\be
	\partial \cdot K = \alpha_S \kappa E \cdot B
\ee
where $\kappa$ is a constant.	This topological charge generates helicity non-conservation.
To understand how this works, consider a parallel electric and magnetic field in electrodynamics
An electron is accelerated and rotates around a magnetic field in the opposite sense to a positron.
Therefore both the electron and positron acquire the same vorticity.  The sign of the vorticity depends upon the sign of $E \cdot B$.    For an extended charge distribution, corresponding to a hadron,
we expect that there will be a similar biasing of the helicity distributions of hadrons.

The classical equations after the collision evolve in time and become dilute as a consequence of the non-linearities of the Yang-Mills equation.  There is a simple solution to this problem which has an
invariance under Lorentz boosts along the collision access.  It has however been recently show that this solution is unstable with respect to small non boost invariant solutions.  These solutions grow in magnitude as time evolves, amplifying small initial fluctuations into full scale chaotically
turbulent solutions.  This turbulence and its rapid onset may be responsible for the early
thermalization seen at RHIC. 

One of the outstanding problems of the Glasma is to understand how these initial fluctuations are formed.  They presumably arise from the initial wavefunction of the nuclei.  Then the classical
instabilities of the Yang-Mills equations amplify these fluctuations, and if one waits long enough, the 
fluctuations dominate the classical solution.  Therefore, quantum noise is amplified to such an effect
that it becomes as large as the classical fields.  Whether or not there is sufficient time
in RHIC or LHC energy collisions for these effects to become significant is not yet known.

\section{The Emerging Picture of RHIC Collisions}

\begin{figure}[ht]
    \begin{center}   
        \includegraphics[width=0.60\textwidth]{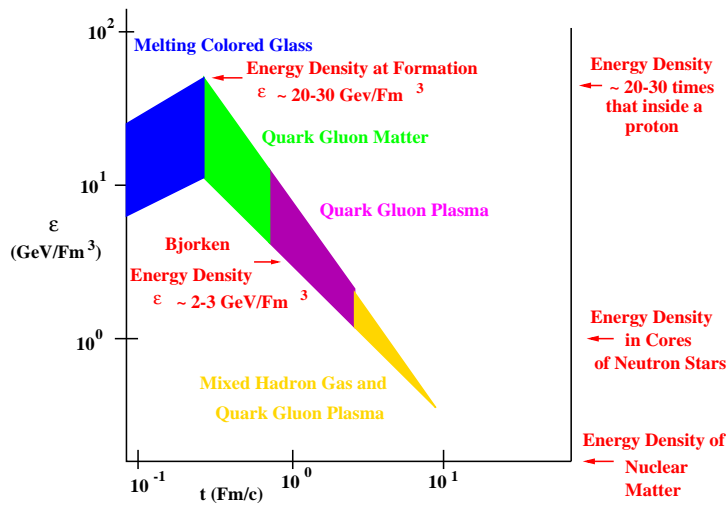}
        \caption{The Emerging Picture of RHIC collisions }
\label{times}
    \end{center}
\end{figure}

The emerging picture we have of RHIC collisions is shown in Fig.  \ref{times}
In the initial state, there are two sheets of colored glass.  They collide and produce a glasma,
which melts into gluons.  During the melting, or perhaps afterwards, the quarks and gluons thermalize.  This eventually makes a Quark Gluon Plasma.  Data from RHIC indicates this happens very rapidly, on a time scale of the order of $1~Fm/c$.  In Fig. \ref{times},  I have presented the typical energy scales and times involved.  The scale of energy comes from the measurements of multiplicities
and HBT radii at RHIC.  The range in estimates comes from making the radical assumptions
of complete thermalization and no thermalization of the gluons.  The bottom line is that the scales are large and the times probed are very early.

\section{Summary}

In addition to the Quark Gluon Plasma, other interesting new forms of matter are being probed at RHIC.
These are the Color Glass Condensate and the Glasma.
These forms of matter allow us to test ideas about QCD when the non-linearities of QCD are present,
yet to use weak coupling methods.  At the LHC, the potential for studying such new form of matter
is great due to the larger range in $x$ and typical momentum scales.

\section{Acknowledgements}
I gratefully acknowledge conversations with 
Francois Gelis,  Edmond Iancu, Dima Kharzeev, and Tuomas Lappi 
Genya Levin, and Raju
Venugopalan on the subject of this talk.  I thank Michael Praszalowicz and Andrzej Bialas for
their kind hospitality at Zakopane.

This manuscript has been authorized under Contract No. DE-AC02-98CH0886 
with
the U. S. Department of Energy.

\end{document}